\documentclass[%
preprint,
superscriptaddress,
preprintnumbers,
nofootinbib,
amsmath,amssymb,
aps,
prd,
]{revtex4-1}
\usepackage[utf8]{inputenc}
\usepackage{bm,amsfonts}
\usepackage{graphicx}
\usepackage{subfigure}
\usepackage{hyperref}

\begin{document}

\title{Primordial black holes and secondary
gravitational waves from k/G inflation}
\author{Jiong Lin}
\email{jionglin@hust.edu.cn}
\affiliation{School of Physics, Huazhong University of Science and Technology, Wuhan, Hubei
430074, China}

\author{Qing Gao}
\email{gaoqing1024@swu.edu.cn}
\affiliation{School of Physical Science and Technology, Southwest University, Chongqing 400715,
China}

\author{Yungui Gong}
\email{Corresponding author. yggong@hust.edu.cn}
\affiliation{School of Physics, Huazhong University of Science and Technology, Wuhan, Hubei
430074, China}

\author{Yizhou Lu}
\email{louischou@hust.edu.cn}
\affiliation{School of Physics, Huazhong University of Science and Technology, Wuhan, Hubei
430074, China}

\author{Chao Zhang}
\email{chaozhang@hust.edu.cn}
\affiliation{School of Physics, Huazhong University of Science and Technology, Wuhan, Hubei
430074, China}

\author{Fengge Zhang}
\email{fenggezhang@hust.edu.cn}
\affiliation{School of Physics, Huazhong University of Science and Technology, Wuhan, Hubei
430074, China}

\begin{abstract}
The possibility that in the mass range around $10^{-12}\ M_\odot$
most of dark matter constitutes of primordial black holes (PBHs)
is a very interesting topic.
To produce PBHs with this mass, the primordial scalar power spectrum needs
to be enhanced to the order of 0.01 at the
scale $k\sim 10^{12}\ \text{Mpc}^{-1}$. The enhanced power spectrum also produces large secondary gravitational waves at the mHz band.
A phenomenological delta function power spectrum
is usually used to discuss the production
of PBHs and secondary gravitational waves. Based on G and k inflations,
we propose a new mechanism to enhance the power spectrum
at small scales by introducing
a non-canonical kinetic term $[1-2G(\phi)]X$ with the function $G(\phi)$ having a peak.
Away from the peak, $G(\phi)$ is negligible and we recover the usual slow-roll
inflation which is constrained by the cosmic microwave background anisotropy observations. Around the peak,
the slow-roll inflation transiently turns to ultra slow-roll inflation.
The enhancement of the power spectrum can be obtained with generic potentials,
and there is no need to fine tune the parameters in $G(\phi)$ to several significant digits.
The energy spectrum $\Omega_{\mathrm{GW}}(f)$ of secondary gravitational waves produced by the model have
the characteristic power law behaviour $\Omega_{\mathrm{GW}}(f)\sim f^{n}$ and is testable
by pulsar timing array and space based gravitational wave detectors.
\end{abstract}

\preprint{2001.05909}

\maketitle

\noindent

\section{Introduction}

The overdense inhomogeneities in the very early universe could gravitationally
collapse to form primordial black holes (PBHs) \cite{Carr:1974nx,Hawking:1971ei}.
PBHs could have a vast range of masses in contrast
to the black hole (BH) formed from the stellar evolution process,
and they were used to explain the
BH binary with tiny effective spin detected by LIGO and Virgo Collaboration \cite{Abbott:2016blz,Abbott:2016nmj, Bird:2016dcv,Sasaki:2016jop}.
Due to the failure of direct detection of particle dark matter (DM),
it is warranted to consider the possibility of PBHs as DM candidate \cite{Ivanov:1994pa, Frampton:2010sw,Belotsky:2014kca,Khlopov:2004sc,
Clesse:2015wea,Carr:2016drx,Inomata:2017okj,Garcia-Bellido:2017fdg,Kovetz:2017rvv}. While light PBHs with mass $M<10^{15}g$ have been evaporated by now
through the Hawking radiation \cite{Hawking:1974sw}, the mass
window for PBHs as DM was strongly constrained by observations
\cite{Gould:1992apjl,Dalcanton:1994apj,Allsman:2000kg,Nemiroff:2001bp,
Wilkinson:2001vv,Tisserand:2006zx,Carr:2009jm,Griest:2013esa,Jacobs:2014yca,
Graham:2015apa,Ali-Haimoud:2016mbv,Wang:2016ana,Carr:2017jsz,
Niikura:2017zjd,Niikura:2019kqi,Laha:2019ssq,Sato-Polito:2019hws,Laha:2020ivk}.
PBHs with mass $10^{15}\text{ g}\lesssim M \lesssim 10^{17}$ g
are constrained by the extragalactic gamma-ray background observations. PBHs with mass $10^{19}\text{ g}\lesssim M \lesssim 10^{20}$ g are constrained by the observations of distribution
of white dwarfs.
PBHs with mass range $1-10^3\ M_\odot$ can be constrained by Pulsar timing array observations.
The microlensing observations in
the Large and Small Magellanic Clouds were used to place constraints on PBHs with mass $10^{26}\text{ g}\lesssim M \lesssim 10^{34}$ g.
The observations of  microlensing events of stars in the Andromeda galaxy (M31) and  Galactic bulge
were used to place constraints on PBHs with mass $10^{22}\text{ g}\lesssim M \lesssim 10^{30}$ g.
The strong lensing of fast radio bursts place constraints on PBHs with mass $M \gtrsim 10\ M_\odot$.
The limit on the merger rate obtained from LIGO/Virgo observations
can be used to constrain PBHs with the mass range $10-300\ M_\odot$.
The observations of cosmic microwave background and Lyman-alpha forests were also be used to constrain the abundance of PBHs.
Therefore, the mass window for PBHs as all dark matter can be around $10^{17}-10^{19}$ g and $10^{20}-10^{22}$ g.

The production mechanism of PBHs could be the direct collapse of primordial
curvature perturbation generated during inflation after horizon reentry
and the formation of PBHs by this mechanism requires the amplitude of
the primordial curvature perturbation $A_s \sim \mathcal{O}(0.01)$ \cite{Sato-Polito:2019hws}. Using the observational constraints on PBH DM and assuming
the piecewise power-law parametrization for the power spectrum
of the primordial curvature perturbation, it was found that
at small scales $k\gtrsim 10^4\ \text{Mpc}^{-1}$ the amplitude of
the power spectrum is $A_s \lesssim 0.05$ \cite{Lu:2019sti}.
However, the amplitude of the primordial curvature perturbation at large scales is
constrained by the cosmic microwave background (CMB) anisotropy measurements to
be $A_s= 2.1 \times 10^{-9}$ at the pivot
scale $k_*=0.05\ \text{Mpc}^{-1}$ \cite{Akrami:2018odb}.
Hence the large enhancement of the amplitude should happen at small scales, but it is impossible to produce
a significant abundance of PBHs as
DM in slow-roll inflationary models by a single canonical scalar field \cite{Motohashi:2017kbs,Passaglia:2018ixg}.
Therefore, to enhance the primordial curvature power spectrum
at small scales, we need to violate the slow-roll condition,
introduce non-canonical kinetic term for the scalar field,
or use more than one scalar field.
The violation of slow-roll condition may be achieved by a field with inflection point
\cite{Germani:2017bcs,Motohashi:2017kbs,Gong:2017qlj}.
Near the inflection point, the velocity of the inflaton
dramatically decreases and the amplitude of the curvature
perturbation is enhanced
\cite{Garcia-Bellido:2017mdw,Ezquiaga:2017fvi,Dimopoulos:2017ged,
Cheng:2018qof,Gao:2018pvq,Cicoli:2018asa,Espinosa:2017sgp,
Espinosa:2018eve,Xu:2019bdp}, but
it is a challenge to fine tune the model parameters to enhance the
amplitude of the primordial curvature perturbation to the order of
$\mathcal{O}(0.01)$ while keeping the total number of e-folds
to be $N\simeq 50-60$ \cite{Sasaki:2018dmp,Passaglia:2018ixg}.
For example, in the critical Higgs inflation and
the axion monodromy inflation, the peak of the power spectrum reaches only to the order $\mathcal{O}(10^{-4})$ \cite{Garcia-Bellido:2017mdw,Ezquiaga:2017fvi,Cheng:2018qof}.
Although the order of $\mathcal{O}(0.01)$ enhancement of
the power spectrum was obtained in \cite{Gong:2017qlj}, but the potential is not a smooth function.

On the other hand, the production of PBHs by the enhanced primordial curvature perturbation
is accompanied by the generation of secondary gravitational waves (GWs)
\cite{Matarrese:1997ay,Mollerach:2003nq,Ananda:2006af,Baumann:2007zm,
Garcia-Bellido:2017aan,Saito:2008jc,Saito:2009jt,Bugaev:2009zh,
Bugaev:2010bb,Alabidi:2012ex,
Orlofsky:2016vbd,Nakama:2016gzw,Inomata:2016rbd,Cheng:2018yyr}.
Therefore, the observations of both PBHs and secondary GWs can be used to constrain
the large enhancement of the amplitude of the primordial curvature perturbation during inflation
and hence to probe the physics in the early universe.
Due to the fine tuning problem mentioned above, a phenomenological delta function \cite{Saito:2008jc,Bartolo:2018evs,Cai:2018dig}, broken power law \cite{Lu:2019sti} or Gaussian power spectrum \cite{Namba:2015gja,Garcia-Bellido:2017aan,Lu:2019sti}
was usually used to discuss the production
of PBHs and secondary GWs \cite{Orlofsky:2016vbd}.
In addition to the large enhancement of the primordial curvature perturbation at small scales,
there are other mechanisms to generate PBHs and GWs at the early universe
\cite{Easther:2006vd,Antusch:2016con,Liu:2017hua,Kamenshchik:2018sig,Chen:2019zza,
Mishra:2019pzq,Fu:2019ttf,Fu:2019vqc,
Liu:2019lul,Cai:2019bmk,DeLuca:2019llr,Vallejo-Pena:2019hgv,Bhattacharya:2019bvk,Ashoorioon:2019xqc},
such as oscillons after inflation \cite{Easther:2006vd,Antusch:2016con,Liu:2017hua}, the double inflation with parametric resonance \cite{Kawasaki:2006zv,Kawasaki:2016pql},
the running-mass model \cite{Stewart:1996ey,Stewart:1997wg,Drees:2011hb}
and the axion-curvaton model \cite{Kasuya:2009up,Kawasaki:2012wr}.

In this paper, based on the expression of the power spectrum in k inflation \cite{ArmendarizPicon:1999rj,Garriga:1999vw}
and G inflation \cite{Kobayashi:2010cm,Kobayashi:2011nu,Kobayashi:2011pc, Herrera:2018ker},
we propose a new mechanism to achieve the order $\mathcal{O}(0.01)$
power spectrum at small scales by introducing a non-canonical
kinetic term $[1-2G(\phi)]X$ with the
function $G(\phi)$ having a peak at $\phi_r$.
The productions of PBHs and secondary GWs are also discussed.
The paper is organized as follows.
In Sec. \ref{section:2}, we review the calculation of the primordial scalar power spectrum first and then propose the enhancement mechanism of the primordial curvature perturbations in the framework of k/G inflation. The PBH abundance and the energy density of secondary GWs generated by this mechanism with a particular model of power law potential are presented in Sec. \ref{section:3}
and Sec. \ref{section:4}, respectively. We conclude the paper in Sec.\ref{section:5}.

\section{k/G inflation}\label{section:2}
The action for G inflation is \cite{Kobayashi:2010cm}
\begin{equation}
\label{eq:action}
S=\int d^4x\sqrt{-g}\left[\frac{1}{2}R+K(\phi,X)-G_{3}(\phi,X)\Box\phi\right].
\end{equation}
where $X=-g_{\mu\nu}\nabla^{\mu}\phi\nabla^{\nu}\phi/2$, $M_{\text{Pl}}=1/\sqrt{8\pi G}=1$, $K$ and $G_3$ are general
functions of $\phi$ and $X$. Assuming that the function $G_{3}$ depends on $\phi$ only,
we can turn the term $G_{3}(\phi)\Box\phi$ to be $-2G_{3\phi}X$ by partial integration,
where $G_{3\phi}=d G_3(\phi)/d\phi$.
Taking the function $K(\phi,X)=X-V(\phi)$,
then the G-inflation model becomes a k-inflation model \cite{ArmendarizPicon:1999rj,Garriga:1999vw},
\begin{equation}
\label{kinfeq}
S=\int d^4x\sqrt{-g}\left[\frac{1}{2}R+X-2G(\phi)X-V(\phi)\right],
\end{equation}
where $G(\phi)=G_{3\phi}$. This can also be thought as general
scalar tensor theory of gravity with non-canonical kinetic term
$\omega(\phi)(\partial\phi)^2$ for the scalar field $\phi$.
Using the spatially flat Friedmann-Robertson-Walker (FRW) metric
and a homogeneous scalar field $\phi=\phi(t)$,
from the action \eqref{kinfeq},
we derive Friedmann equations
\begin{gather}
\label{Eq:eom1}
3H^2=\frac{1}{2}\dot{\phi}^2+V(\phi)-\dot{\phi}^2G(\phi),\\
\label{Eq:eom2}
2\dot{H}+3H^2+\frac{1}{2}\dot{\phi}^2-V(\phi)-\dot{\phi}^2G(\phi)=0,\\
\label{Eq:eom3}
\ddot{\phi}+3H\dot{\phi}+\frac{V_{\phi}-\dot{\phi}^2G_{\phi}}{1-2G(\phi)}=0,
\end{gather}
where $G_\phi=dG(\phi)/d\phi$.
We define the slow-roll parameters
\begin{equation}
\epsilon_1=-\frac{\dot{H}}{H^2},\ \epsilon_2=-\frac{\ddot{\phi}}{H\dot{\phi}},\
 \epsilon_3=\frac{G_{\phi}\dot{\phi}^2}{V_{\phi}},
\end{equation}
so slow-roll inflation is realized when $|\epsilon_i|\ll1$, where $i=1,2,3$.
By using Eqs. \eqref{Eq:eom1} and \eqref{Eq:eom2},
the first slow-roll parameter $\epsilon_1$ can be expressed as
\begin{equation}
\label{eps1:sr1}
\epsilon_1=\frac{X(1-2G)}{H^2}.
\end{equation}
Under slow-roll approximation, Eqs. \eqref{Eq:eom1} and  \eqref{Eq:eom3} can be expressed as
\begin{gather}
  3H^2\simeq V, \\
  3H\dot{\phi}(1-2G)+V_{\phi}\simeq0.
\end{gather}
To the first order of approximation, the quadratic action for the curvature perturbation $\zeta$ is \cite{Kobayashi:2010cm,Garriga:1999vw},
\begin{equation}
  S^{(2)}=\frac{1}{2}\int d\tau d^3x\tilde{z}^2[\mathcal{G}(\zeta')^2-\mathcal{F}(\vec{\nabla}\zeta)^2],\label{eq:cp}
\end{equation}
where $\tilde{z}=a\dot{\phi}/H$,
$\mathcal{F}=\mathcal{G}=1-2G$,
and the prime represents derivative with respect to the conformal time $\tau$.
Since the sound speed for the scalar mode is $c_s^2=\mathcal{F}/\mathcal{G}=1$,
so there is no problem with ghost and gradient instabilities.
By varying the quadratic action Eq. \eqref{eq:cp} with the respect to
the curvature perturbation $\zeta_k$ in the Fourier space, we get
\begin{equation}
\label{eq:1}
u''_k+\Bigl(k^2-\frac{z''}{z}\Bigl)u_k=0,
\end{equation}
where $z=(1-2G)^{1/2}\tilde{z}$ and $u_k=z\zeta_{k}$.
Solving the mode equation \eqref{eq:1}, we obtain the scalar power spectrum
\begin{equation}
\label{eq:ps}
P_{\zeta}=\frac{H^4}{4\pi^2\dot{\phi}^2(1-2G)}\simeq\frac{V^3}{12\pi^2V_{\phi}^2}(1-2G),\\
\end{equation}
and the scalar spectral index
\begin{equation}
\label{nseq1}
n_s-1=\frac{1}{1-2G}\left(2\eta_V-6\epsilon_V+
\frac{2G_{\phi}}{1-2G}\sqrt{2\epsilon_V}\right),
\end{equation}
where $\epsilon_V=(V'/V)^2/2$ and $\eta_V=V''/V$.
The scalar field does not affect the tensor perturbation,
so the tensor power spectrum is \cite{Kobayashi:2010cm,Garriga:1999vw}
\begin{equation}
\label{ptspec}
P_T=\frac{H^2}{2\pi^2},
\end{equation}
and the tensor to scalar ratio reads
\begin{equation}
\label{req1}
r=\frac{P_T}{P_{\zeta}}=\frac{16 X(1-2G)}{H^2}=16\epsilon_1
\simeq 16\frac{\epsilon_V}{1-2G}.
\end{equation}


From the expression \eqref{eq:ps} for the scalar power spectrum,
we propose a mechanism to enhance the power spectrum by a choosing
suitable function $G(\phi)$. At large scales (40-60 e-folds before the end of inflation), if $G(\phi)\approx 0$, then the effect
of $G(\phi)$ is negligible and the results from slow-roll inflation
are not changed, so the observational constraints can be satisfied.
At small scales, if the function $1-2G(\phi)$ has a peak, then
the enhancement of the power spectrum is achieved. Therefore,
the enhancement we proposed requires that the function $G(\phi)$ has
a peak and away from the peak it decays to zero.
To be specific, we choose the function
\begin{equation}
\label{gfuneq1}
-2G_a(\phi)=\frac{d}{1+\left|\frac{\phi-\phi_{r}}{c}\right|},
\end{equation}
where the parameters $\phi_{r}$ and $c$ have the dimension of mass,
$c$ controls the width of the peak and
the dimensionless parameter $d$ determines the height of the peak.
For Brans-Dicke theory in Jordan frame,
the non-canonical kinetic term is $X/\phi$ \cite{Brans:1961sx}
with the choice of $\omega(\phi)=1/\phi$.
If we make a shift $\phi\to \phi+a$, then the kinetic term becomes $X/(a+\phi)$.
Note that other choices of the peak function $G(\phi)$ are also possible to realize the mechanism.
For comparison, we also choose
\begin{equation}
\label{gfuneq2}
-2G_b(\phi)=\frac{d}{\sqrt{1+(\frac{\phi-\phi_{r}}{c})^2}}.
\end{equation}
Away from the peak, $|\phi-\phi_r|\gg c$, the function $G_b(\phi)$ decays as $\sim 1/\phi$.
Around the peak, $|\phi-\phi_r|\ll c$,  the function $G_b(\phi)$ approaches the peak quicker than $G_a(\phi)$ because
\begin{equation}
\label{gbsim}
G_b(\phi)\approx d\left[1-\frac{1}{2}\left(\frac{\phi-\phi_r}{c}\right)^2\right].
\end{equation}
The square root potential $\sqrt{a+b\phi^2}$ can be obtained by
a D5 brane wrapped on a two-cycle in axion monodromy \cite{Silverstein:2008sg,McAllister:2008hb}.
For the nonminimal coupling $f(\phi)R$, we have non-canonical
kinetic terms after a conformal transformation. In particular,
for the nonminimal coupling $-\phi^2 R/6$, in Einstein frame the kinetic term becomes $X/(1-\phi^2/6)^2$ which has a peak at $\phi=\sqrt{6}$ \cite{Kallosh:2013hoa}. More generally, the kinetic term for superconformal
$\alpha$-attractors is $X/[1-\phi^2/(6\alpha)]^2$ \cite{Kallosh:2013yoa}.
These arguments motivate the form of the phenomenological functions \eqref{gfuneq1} and \eqref{gfuneq2}
although we are not sure how to derive them from a first principle.

To get the enhancement over seven orders of magnitude, $d$ should be in the order of $10^8$. In this paper, we choose $d=5.26\times 10^8$. Note that this choice is arbitrary.
The number of e-folds around the peak is
\begin{equation}
\label{nefolds}
\Delta N=\int^{\phi_{r}+\Delta\phi}_{\phi_{r}-\Delta\phi}\frac{H}{\dot{\phi}}d\phi
\simeq-\int^{\phi_{r}+\Delta\phi}_{\phi_{r}-\Delta\phi}\frac{V(1-2G)}{V_{\phi}}d\phi.\\
\end{equation}
Apparently, the number of e-folds around the peak can be very large.
To keep the total number of e-folds before the end of inflation
to be $N\simeq 50-60$,
the peak width $c$ should be very small.

Due to the peak in $G(\phi)$, we may worry about the exit of inflation
around $\phi_r$ because of Eq. \eqref{eps1:sr1}. However, Eq. \eqref{Eq:eom3}
tells us that $\dot\phi$ decreases dramatically around the peak.
Therefore, it behaves like ultra slow-roll inflation \cite{Tsamis:2003px,Kinney:2005vj,Yi:2017mxs} around $\phi_r$.
Note that because $\dot{\phi}^2G_{\phi}$ may dominate over $V_{\phi}$, the effective potential may $(V_{\phi}-\dot{\phi}^2G_{\phi})/(1-2G(\phi))<0$ and thus
$\epsilon_2>3$.

\begin{figure}[htp]
\centering
\includegraphics[width=0.7\textwidth]{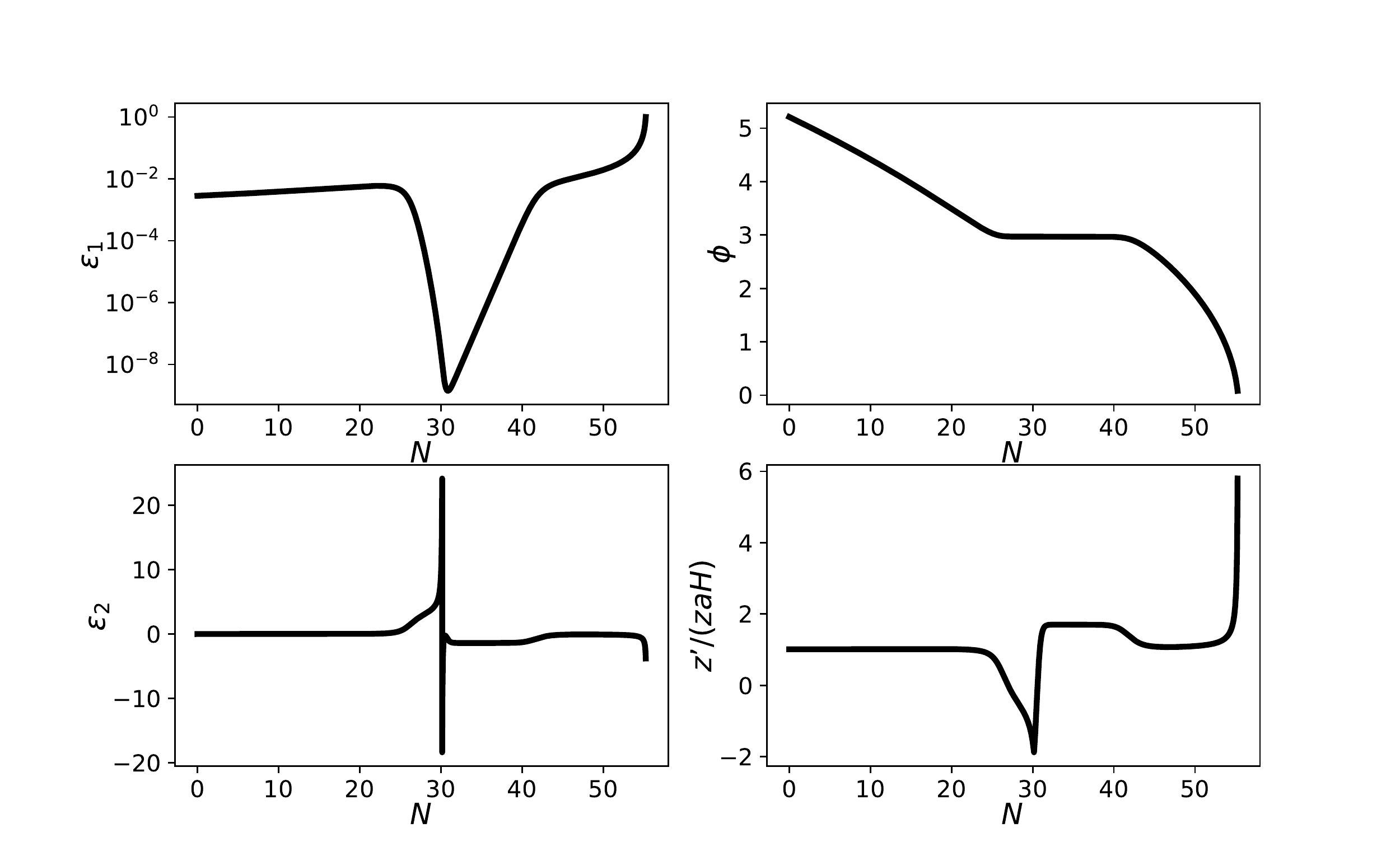}
\caption{The evolution of $\phi$, $\epsilon_1$, $\epsilon_2$ and
$z'/(zaH)$ with the parameter set D and  the peak function $G_a(\phi)$. We take the number of e-folds $N$
before the end of inflation as the time.}
\label{Fig:1}
\end{figure}

Because of the violation of slow-roll conditions, the expression \eqref{eq:ps}
for the scalar power spectrum may not be applied and the enhancement of
the power spectrum may not reached.
Let us exam how the power spectrum could be enhanced without
using the formula \eqref{eq:ps}.
The Fourier component of the curvature perturbation satisfies
\begin{equation}
\label{eq:f}
\zeta''_k+2\frac{z'}{z}\zeta'_k+k^2\zeta_k=0,
\end{equation}
where
\begin{equation}
\frac{z'}{z}=aH\left[1+\epsilon_1-\epsilon_2-\frac{G_{\phi}\dot{\phi}}{H(1-2G)}\right].
\end{equation}
When the velocity of the scalar field dramatically decreases and $\epsilon_2>3$, the friction term in Eq. \eqref{eq:f} transiently changes sign, i.e., $z'/z<0$, as shown in Fig. \ref{Fig:1}. The friction term becomes a driving term and hence the curvature perturbation $\zeta_k$ increases and the power spectrum is enhanced in this regime.


To show how the model and the above enhancement mechanism work,
we consider the power law potential $V(\phi)=\lambda\phi^p$ as an example.
To be consistent with the observational constraint by CMB measurements \cite{Akrami:2018odb},
we choose $p=2/5$ and $\lambda=7.2\times10^{-10}$.
At the CMB scale $k_*=0.05\ \text{Mpc}^{-1}$ when the mode exits the horizon,
the field value is $\phi_*=5.21$, the scalar spectral tilt $n_s \simeq 0.97$,
the tensor to scalar ratio $r\simeq 0.045$, the amplitude of the power spectrum
is $A_s\simeq 2.1\times 10^{-9}$ and the number of e-folds before the end
of inflation is $N \simeq 54$.

To get large enhancement $O(0.01)$ at small scales,
we choose the parameters in functions $G_a(\phi)$ and $G_b(\phi)$ as listed in Tables \ref{tab:1} and \ref{tab:2}.
As shown in Table \ref{tab:1}, the enhancement scale is adjusted by the parameter $\phi_r$.
If $\phi_r$ is further away from $\phi_*$, then the enhancement scale becomes smaller.
As discussed above, the peak in $G(\phi)$ violates the slow-roll condition,
so we numerically solve Eq. \eqref{eq:f} to obtain the power spectrum and
the corresponding values at the peak scales as shown in Table \ref{tab:1} for $G_a(\phi)$ and \ref{tab:2} for $G_b(\phi)$.
These results show that both functions $G_a(\phi)$ and $G_b(\phi)$
work and they give similar results, so we present
the detailed results for
the function $G_a(\phi)$ only in the following discussion.

In Fig. \ref{Fig:1} we plot the evolutions of $\phi$, $\epsilon_1$, $\epsilon_2$ and
$z'/(zaH)$ with the parameter set D and the function $G_a(\phi)$.
We see that around $N \sim 30$
the inflaton rolls very slowly and its velocity decreases to be very small,
so the slow-roll parameter $\epsilon_1$ becomes negligible which enhances the power
spectrum dramatically.
Note that the slow-roll parameter $\epsilon_2$ changes quickly and becomes very
large.
It indicates the effective potential transiently changes sign, i.e., $(V_{\phi}-\dot{\phi}^2G_{\phi})/(1-2G(\phi))<0$ and $\dot{\phi}$ dramatically decreases.

\begin{table*}[htp]
  \centering
  \caption{The chosen parameter sets and the results for the scalar power spectrum at small peak scales, PBH abundances and critical frequency of secondary gravitational waves with the peak function $G_a(\phi)$.}
  \begin{tabular}{cccccccccccc}
  \hline
  Sets &$\phi_{r}$&$c$&$n_s$&$k_{\text{peak}}/\text{Mpc}^{-1}$&$P_{\zeta(\text{peak})}$&$M_{\text{pbh}}^{\text{peak}}/M_{\odot}$&$Y_{\text{PBH}}^{\text{peak}}$&$f_c$/Hz\\
  \hline
  A &4.5&$9.54\times10^{-11}$&0.9736&$2.86\times10^5$&$1.66\times10^{-2}$&$28.9$&$7.7\times10^{-5}$&$4.43\times10^{-10}$\\
  B &4.5&$9.568\times10^{-11}$&0.9737&$2.7\times10^5$&$1.86\times10^{-2}$&32.5&$0.001$&$4.18\times10^{-10}$\\
  C &4.1&$1.05\times10^{-10}$&0.969&$3\times10^7$&$1.49\times10^{-2}$&$0.0026$&$4.7\times10^{-4}$&$4.6\times10^{-8}$\\
  D &2.97&$1.472\times10^{-10}$&0.967&$1.63\times10^{12}$&$1.32\times10^{-2}$&$9\times10^{-13}$&$0.73$&$2.5\times10^{-3}$\\
  \hline
  \end{tabular}
\label{tab:1}
\end{table*}

\begin{table*}[htp]
  \centering
  \caption{The chosen parameter sets and the results for the scalar power spectrum at small peak scales, PBH abundances and critical frequency of secondary gravitational waves with the peak function $G_b(\phi)$.}
  \begin{tabular}{ccccccccccc}
  \hline
  Sets &$\phi_{r}$&$c$&$n_s$&$k_{\text{peak}}/\text{Mpc}^{-1}$&$P_{\zeta(\text{peak})}$&$M_{\text{pbh}}^{\text{peak}}/M_{\odot}$&$Y_{\text{PBH}}^{\text{peak}}$&$f_c$/Hz\\
  \hline
  A &4.5&$9.54\times10^{-11}$&0.9736&$2.97\times10^5$&$1.88\times10^{-2}$&26.8&$0.00148$&$4.6\times10^{-10}$\\
  C &4.1&$1.05\times10^{-10}$&0.969&$3.16\times10^7$&$1.7\times10^{-2}$&$0.002$&$0.0167$&$4.9\times10^{-8}$\\
  E &2.97&$1.4658\times10^{-10}$&0.967&$1.36\times10^{12}$&$1.33\times10^{-2}$&$1.28\times10^{-12}$&$0.85$&$2.1\times10^{-3}$\\
  \hline
  \end{tabular}
\label{tab:2}
\end{table*}

\begin{figure}[htp]
\centering
\includegraphics[width=0.7\textwidth]{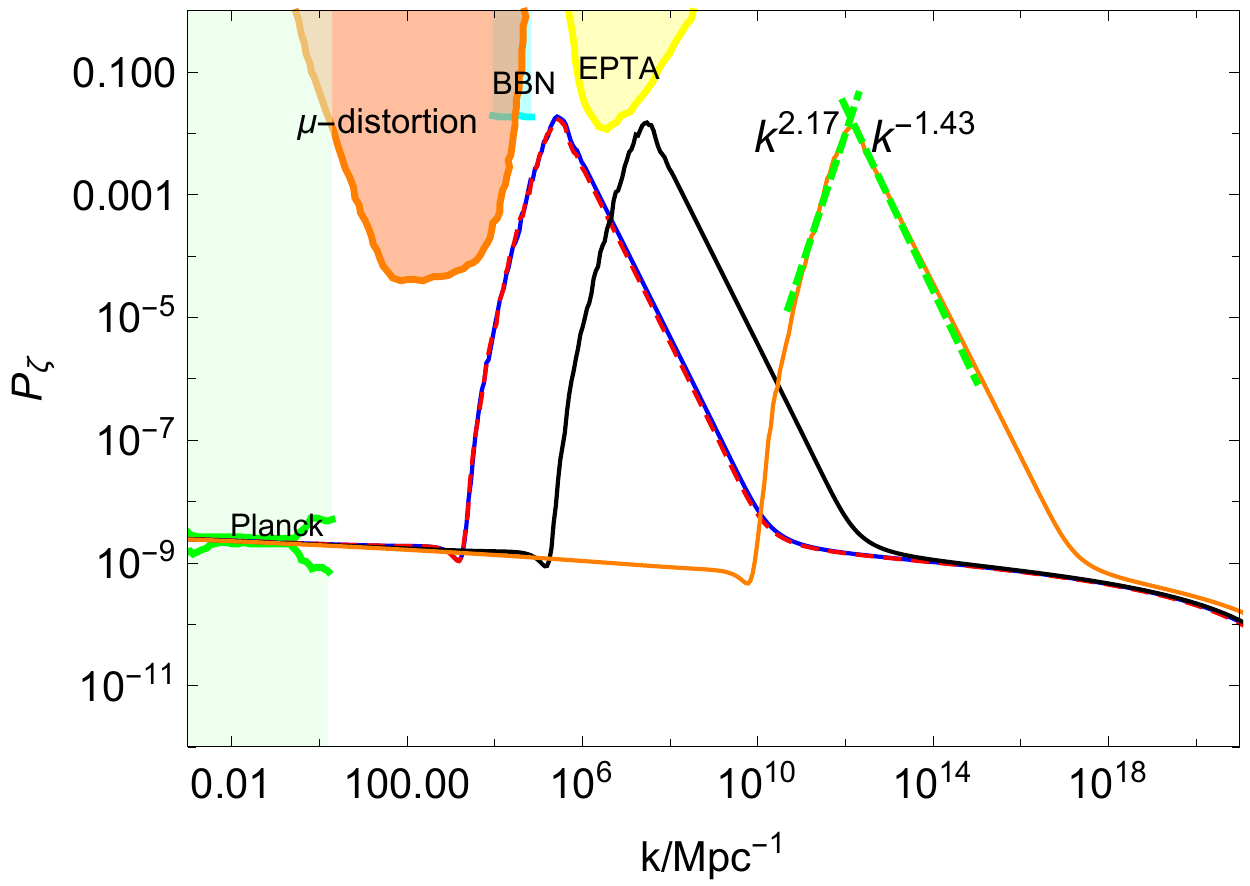}
\caption{The results for the scalar power spectrum  with the peak function $G_a(\phi)$. The blue line uses the parameter
set A, the red dashed line uses the parameter set B, the black line uses the parameter set C and the oranges line uses the parameter set D.
The dashed green lines show the scale dependent behaviour of the power spectrum.
The lightgreen shaded region is excluded by the CMB observations \cite{Akrami:2018odb}. The yellow, cyan and orange regions show the
constraints from the PTA observations \cite{Inomata:2018epa},
the effect on the ratio between neutron and proton
during the big bang nucleosynthesis (BBN) \cite{Inomata:2016uip}
and $\mu$-distortion of CMB \cite{Fixsen:1996nj}, respectively.
}
\label{Fig:2}
\end{figure}

In Fig. \ref{Fig:2}, we show the results for the scalar power spectrum generated with the function $G_a(\phi)$.
At large scales, the power spectrum is in the order of $\mathcal{O}(10^{-9})$, which is compatible with CMB constraints \cite{Akrami:2018odb}.
At small scales, the power spectrum is enhanced to the
order of $\mathcal{O}(0.01)$, which is large enough to
produce PBHs after the horizon reentry as discussed below.
It is interesting to note that the power spectrum can be parameterized as the broken
power law form $P_{\zeta}\sim k^n$. For the parameter set D, $P_{\zeta}\sim k^{2.17}$ for $k<k_c=1.63\times10^{12}\ \text{Mpc}^{-1}$ and $P_{\zeta}\sim k^{-1.43}$ for $k>k_c=1.63\times10^{12}\ \text{Mpc}^{-1}$.
The models also satisfy the constraints from CMB $\mu$-distortion,
big bang nucleosynthesis (BBN) and pulsar timing array (PTA) observations \cite{Inomata:2018epa,Inomata:2016uip,Fixsen:1996nj}.

\section{PBH abundance}\label{section:3}
When the primordial curvature perturbation reenters the horizon during radiation dominated era, it may gravitationally collapse to form PBHs. The PBH mass is equal to $\gamma M_{\mathrm{hor}}$, where $M_{\mathrm{hor}}$ is the horizon mass and we choose the factor
$\gamma= 0.2$ \cite{Carr:1975qj}. The current fractional energy density of PBHs with mass $M$ to DM is \cite{Carr:2016drx,Gong:2017qlj}
\begin{equation}
\label{fpbheq1}
\begin{split}
Y_{\text{PBH}}(M)=&\frac{\beta(M)}{3.94\times10^{-9}}\left(\frac{\gamma}{0.2}\right)^{1/2}
\left(\frac{g_*}{10.75}\right)^{-1/4}\\
&\times \left(\frac{0.12}{\Omega_{\text{DM}}h^2}\right)
\left(\frac{M}{M_\odot}\right)^{-1/2},
\end{split}
\end{equation}
where $M_{\odot}$ is the solar mass, $g_*$ is the effective degrees of freedom at the formation time, $\Omega_{\text{DM}}$ is the current
energy density parameter of DM,
the fractional energy density of PBHs
at the formation is \cite{Young:2014ana, Ozsoy:2018flq,Tada:2019amh}
\begin{equation}
\label{eq:beta}
\beta(M)\approx\sqrt{\frac{2}{\pi}}\frac{\sigma(M)}{\delta_c}
\exp\left(-\frac{\delta_c^2}{2\sigma^2(M)}\right),
\end{equation}
$\delta_c$ is the critical density perturbation for the PBH formation,
$\sigma(k)$ is the mass variance associated with the PBH mass $M(k)$
smoothing on the comoving horizon length $k^{-1}=1/(aH)$ \cite{Ozsoy:2018flq,Young:2014ana}
\begin{equation}
\label{sigmaeq1}
\sigma^2(k)=\left(\frac{4}{9}\right)^2\int \frac{dq}{q} W^2(q/k)(q/k)^4P_{\zeta}(q),
\end{equation}
and the Gaussian window function $W(x)=\exp(-x^2/2)$.
The effective degrees of freedom $g_*=107.5$ for $T>300$GeV
and $g_*=10.75$ for $0.5\text{MeV}<T<300\text{GeV}$.
We take the observational value $\Omega_{\text{DM}}h^2=0.12$ \cite{Aghanim:2018eyx}
and $\delta_c=0.4$ \cite{Musco:2012au,Harada:2013epa,Tada:2019amh,Escriva:2019phb,Yoo:2020lmg}
for the calculation of PBH abundance.
The relation between the PBH mass $M$ and the scale $k$ is \cite{Gong:2017qlj}
\begin{equation}
\label{mkeq1}
M(k)=3.68\left(\frac{\gamma}{0.2}\right)\left(\frac{g_*}{10.75}\right)^{-1/6}
\left(\frac{k}{10^6\ \text{Mpc}^{-1}}\right)^{-2} M_{\odot}.
\end{equation}
With the approximation that the power spectrum is scale invariant, we get $\sigma(k)\simeq (4/9)\sqrt{P_{\zeta}}$ and
$$\beta(M) \approx \sqrt{\frac{2}{\pi}}\frac{\sqrt{P_{\zeta}}}{\mu_c}
\exp\left(-\frac{\mu_c^2}{2P_{\zeta}}\right),$$ where $\mu_c=9\delta_c/4$.

\begin{figure}[htp]
\centering
\includegraphics[width=0.7\textwidth]{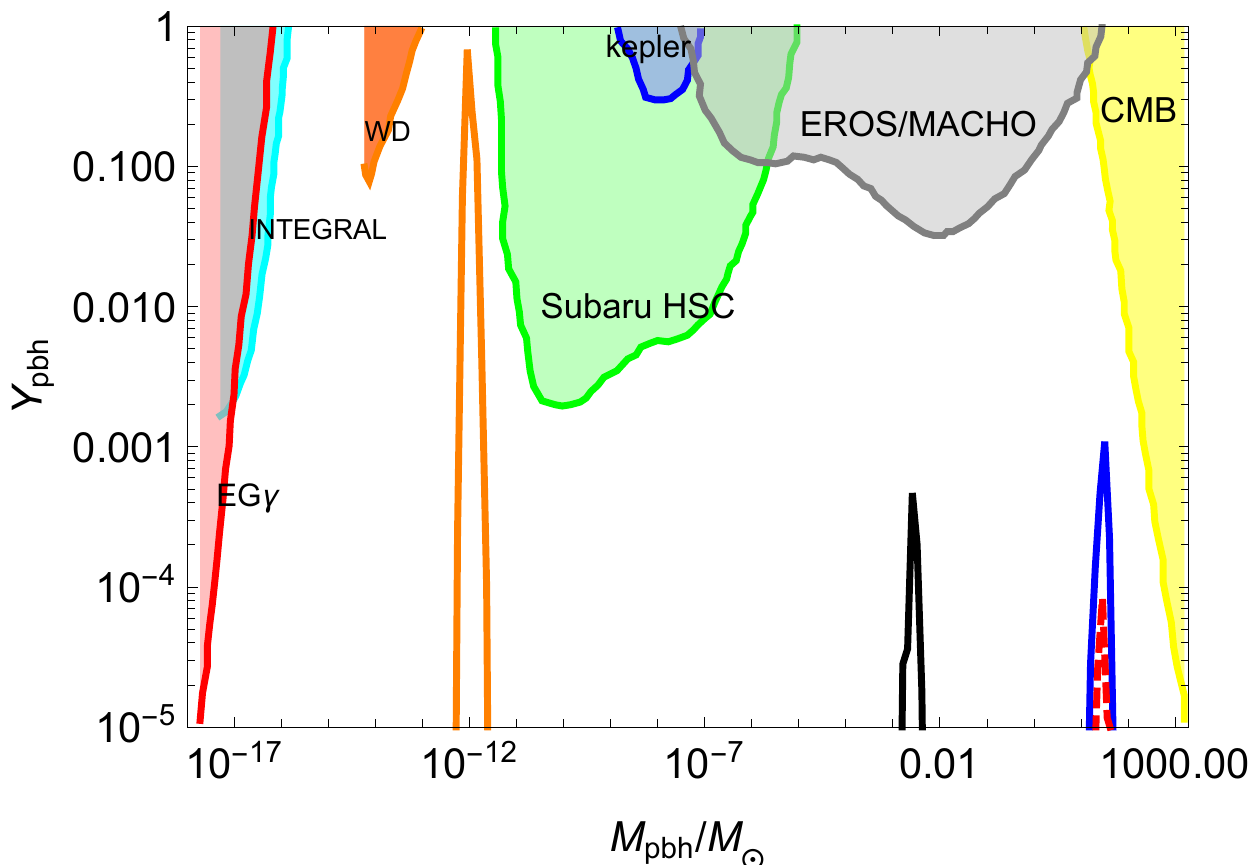}
\caption{The PBH abundances for the parameter sets A (the blue line), the parameter sets B (the red dashed line),
C (the black line) and D (the orange line)  with the peak function $G_a(\phi)$.
The shaded regions show the observational constraints on the PBH abundance:
the yellow region from accretion constraints by CMB \cite{Ali-Haimoud:2016mbv,Poulin:2017bwe},
the red region from extragalactic gamma-rays by PBH evaporation (EG$\gamma$) \cite{Carr:2009jm}, the cyan region from galactic center 511 keV gamma-ray line (INTEGRAL) \cite{Laha:2019ssq,Dasgupta:2019cae}, the orange region from white dwarf explosion (WD) \cite{Graham:2015apa},
the green region from microlensing events with Subaru HSC \cite{Niikura:2017zjd},
the blue region from the Kepler satellite \cite{Griest:2013esa},
the gray region from the EROS/MACHO \cite{Tisserand:2006zx}.}
\label{Fig:3}
\end{figure}

Substituting the obtained power spectrum into Eqs. \eqref{fpbheq1}, \eqref{eq:beta},
\eqref{sigmaeq1} and \eqref{mkeq1}, we get the PBH abundances as shown in
Table \ref{tab:1} and Fig. \ref{Fig:3}.
For the parameter sets A and B, the model produces PBHs
with mass $M \simeq 30 M_{\odot}$ and abundance $Y^{\text{peak}} \simeq 7.7\times10^{-5}$ for parameter set A and
$Y^{\text{peak}} \simeq 0.001$ for parameter set B,
which may explain the BH event GW150914 observed by LIGO \cite{Abbott:2016blz}.
Although the peak of the power spectrum from the parameter set A
is only 11\% smaller than that from the parameter set B, but
the produced PBH abundance is almost two orders smaller.
In this mass range PBHs cannot consist of all DM due to the constraints from CMB \cite{Ali-Haimoud:2016mbv}.
For the parameter set C, the model produces PBHs
with mass $M \simeq 2.6\times 10^{-3} M_{\odot}$ and abundance $Y^{\text{peak}} \simeq 4.7\times10^{-4}$.
For the parameter set D, the model produces PBHs with mass $M \simeq 9\times 10^{-13}M_{\odot}$. The produced PBH abundance is $Y^{\text{peak}}\simeq 0.73$.
In this mass range, the observational constraint on PBH abundances
are absent \cite{Nakamura:1997sm}, so all DM can be PBHs.

\section{Secondary GWs}\label{section:4}
\begin{figure}[htp]
\centering
\includegraphics[width=0.7\textwidth]{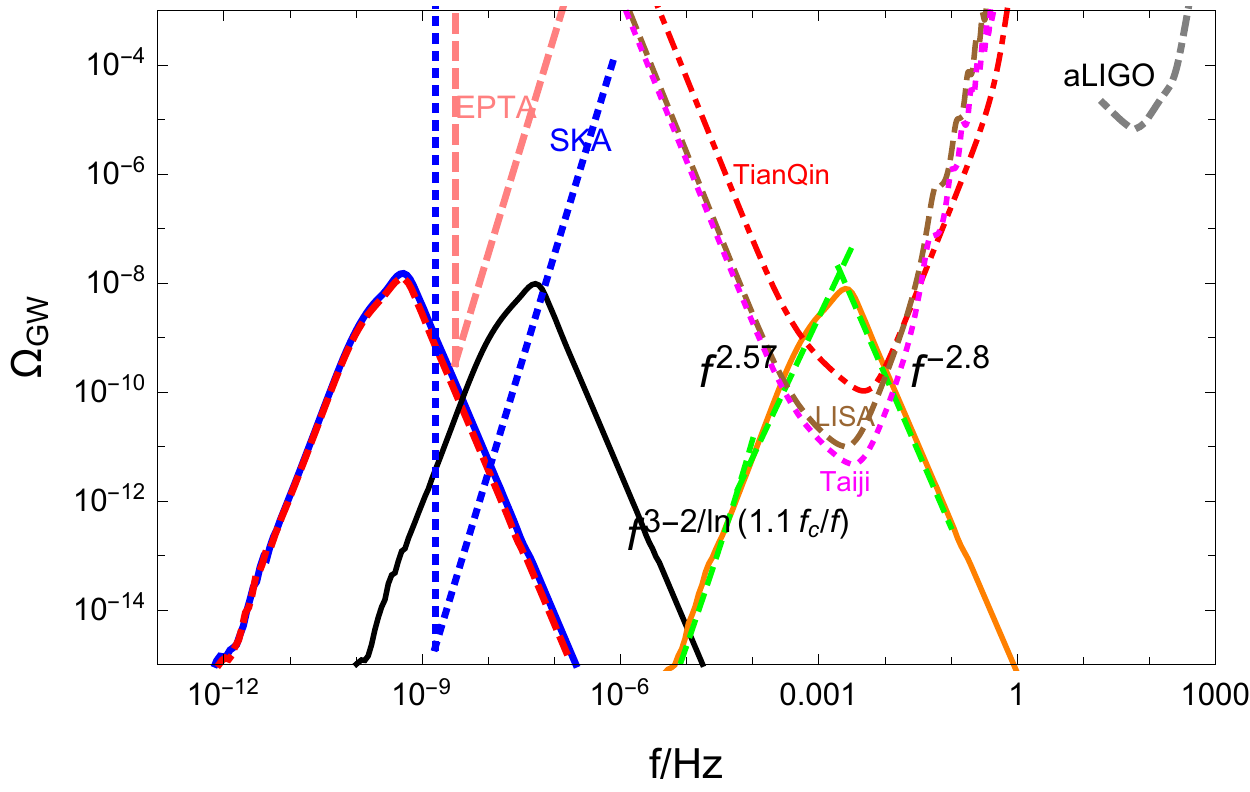}
\caption{The secondary GWs generated in the model with the peak function $G_a(\phi)$.
The solid blue, dashed red line, solid black and solid orange denote
the results for the model parameter sets A, B, C and D, respectively.
The dashed green lines show the broken power law behaviour of $\Omega_{\text{GW}}$.
The pink dashed curve denotes the EPTA limit \cite{Ferdman:2010xq,Hobbs:2009yy,McLaughlin:2013ira,Hobbs:2013aka} ,
the blue dotted curve denotes the SKA limit \cite{Moore:2014lga},
the red dot-dashed curve in the middle denotes the TianQin limit \cite{Luo:2015ght},
the dotted magenta curve shows the TaiJi limit \cite{Hu:2017mde},
the brown dashed curve shows the LISA limit \cite{Audley:2017drz},
and the gray dotdashed curve denotes the aLIGO limit \cite{Harry:2010zz,TheLIGOScientific:2014jea}.}
\label{Fig:4}
\end{figure}

In addition to the production of PBHs,
the large density perturbations generated at small scales during inflation
could produce secondary GWs
and be tested by the future PTA observations and space based GW observatory.
The Fourier components of the second order tensor perturbations $h_{\bm{k}}$ satisfy
\cite{Ananda:2006af,Baumann:2007zm}
\begin{equation}
\label{eq:hk}
h''_{\bm{k}}+2\mathcal{H}h'_{\bm{k}}+k^2h_{\bm{k}}=4S_{\bm{k}},
\end{equation}
with the scalar source
\begin{equation}
\label{hksource}
\begin{split}
S_{\bm{k}}=\int \frac{d^3\tilde{k}}{(2\pi)^{3/2}}e_{ij}(\bm{k})\tilde{k}^i\tilde{k}^j
\left[2\Phi_{\tilde{\bm{k}}}\Phi_{\bm{k}-\tilde{\bm{k}}}
+\frac{4}{3(1+\omega)\mathcal{H}^2} \right.\\
\left. \times\left(\Phi'_{\tilde{\bm{k}}}+\mathcal{H}\Phi_{\tilde{\bm{k}}}\right)
\left(\Phi'_{\bm{k}-\tilde{\bm{k}}}+\mathcal{H}\Phi_{\bm{k}-\tilde{\bm{k}}}\right)\right],
\end{split}
\end{equation}
where $\mathcal{H}= aH$, $\omega=p/\rho$, $e_{ij}(\bm{k})$ is the polarization tensor, the Bardeen potential $\Phi_{\bm{k}}=\Psi(k\eta)\phi_{\bm{k}} $,
the transfer function $\Psi$ in the radiation domination is
\begin{equation}
\label{transfer}
\Psi(x)=\frac{9}{x^2}\left(\frac{\sin(x/\sqrt{3})}{x/\sqrt{3}}-\cos(x/\sqrt{3})\right),
\end{equation}
and the primordial value $\phi_{\bm{k}}$ is
\begin{equation}
\label{phikeq4}
\langle\phi_{\bm{k}}\phi_{\tilde{\bm{k}}}\rangle
=\delta^{(3)}(\bm{k}+\tilde{\bm{k}})\frac{2\pi^2}{k^3}\left(\frac{3+3w}{5+3w}\right)^2 P_\zeta(k),
\end{equation}
The power spectrum of the induced GWs is defined as
\begin{equation}
\label{eq:pwrh}
\langle h_{\bm{k}}(\eta)h_{\tilde{\bm{k}}}(\eta)\rangle
=\frac{2\pi^2}{k^3}\delta^{(3)}(\bm{k}+\tilde{\bm{k}})P_h(k,\eta),
\end{equation}
The Green's function for Eq. \eqref{eq:hk} is
\begin{equation}
\label{greenfunc}
g_k(\eta,\eta')=\frac{\sin[k(\eta-\eta')]}{k}.
\end{equation}
Solving Eq. \eqref{eq:hk} by using the Green function method with the Green's function \eqref{greenfunc},
we obtain the power spectrum of the induced GWs \cite{Baumann:2007zm,Ananda:2006af}
\begin{equation}
\begin{split}
P_h(k,\eta)=
4\int_{0}^{\infty}dv\int_{|1-v|}^{1+v}du
\left\{\left[\frac{4v^2-(1-u^2+v^2)^2}{4uv}\right]^2\right.\\
\left.\times I_{RD}^2(u,v,x\to \infty)P_{\zeta}(kv)P_{\zeta}(ku)\right\},
\end{split}
\end{equation}
where $u=|\bm{k}-\tilde{\bm{k}}|/k$, $v=\tilde{k}/k$, $x=k\eta$
and the integral kernel $I_{\text{RD}}$ is \cite{Espinosa:2018eve,Lu:2019sti}
\begin{equation}
\label{irdeq1}
\begin{split}
I_{\text{RD}}=&\int_1^x dy\, y \sin(x-y)\{3\Psi(uy)\Psi(vy)\\
&+y[\Psi(vy)u\Psi'(uy)+v\Psi'(vy)\Psi(uy)]\\
&+y^2 u v \Psi'(uy)\Psi'(vy)\},
\end{split}
\end{equation}
and an analytical expression for $I_{\text{RD}}$ was given in Refs. \cite{Espinosa:2018eve,Lu:2019sti}.
The energy density of induced GWs generated
in the radiation domination is \cite{Inomata:2016rbd,Kohri:2018awv}
\begin{equation}
\label{gwres1}
\begin{split}
\Omega_{\mathrm{GW}}(k,\eta)=&\frac{1}{6}\left(\frac{k}{aH}\right)^2
\int_{0}^{\infty}dv\int_{|1-v|}^{1+v}du\left\{ \right.\\
&\left[\frac{4v^2-(1-u^2+v^2)^2}{4uv}\right]^2\\
&\left. \times \overline{I_{\text{RD}}^{2}(u, v, x\to \infty)} P_{\zeta}(kv)P_{\zeta}(ku)\right\},
\end{split}
\end{equation}
where $\overline{I_{\text{RD}}^{2}}$ is the oscillation time average.
Since GWs behave like radiation, the current energy densities of GWs are related to their values well
after the horizon reentry in the radiation dominated era
\begin{equation}\label{gwres2}
\Omega_{GW}(k,\eta_0)=\Omega_{GW}(k,\eta)\frac{\Omega_{r}(\eta_0)}
{\Omega_{r}(\eta)},
\end{equation}
where $\Omega_r$ is the fraction energy density of radiation.
Plugging the power spectrum in Fig. \ref{Fig:2} into Eq. \eqref{gwres1}\eqref{gwres2}
and using Eqs. \eqref{transfer} and \eqref{irdeq1} we obtain current energy densities of the induced
GWs and the results are shown in Fig. \ref{Fig:4}.
If we use the analytical expression for $I_{\text{RD}}$
in Ref. \cite{Kohri:2018awv}, the difference on the secondary
GWs is small for the power spectrum discussed in this paper.
To compare the results with observations, in Fig. \ref{Fig:4},
we also show the sensitivity curves for European PTA (EPTA) \cite{Ferdman:2010xq,Hobbs:2009yy,McLaughlin:2013ira,Hobbs:2013aka}, the Square Kilometer Array (SKA) \cite{Moore:2014lga},
Advanced Laser Interferometer Gravitational Wave Observatory (aLIGO) \cite{Harry:2010zz,TheLIGOScientific:2014jea},
Laser Interferometer Space Antenna (LISA) \cite{Danzmann:1997hm,Audley:2017drz}, TaiJi \cite{Hu:2017mde} and TianQin \cite{Luo:2015ght}.

As shown in Fig. \ref{Fig:4}, for the parameter set D,
the induced GWs are in the mHz band and could
be tested by the future space based detector like
LISA, TaiJi and TianQin. The induced GWs
from the parameter sets A and B have
the peak frequency $f\sim 10^{-10}\text{Hz}$ and those from the parameter set  C
have the peak frequency $f\sim10^{-8}\text{Hz}$,
both of them could be tested by SKA.
The signal to noise ratio (SNR) is  \cite{Thrane:2013oya,Smith:2019wny}
\begin{equation}
\text{SNR}^2=T\int_0^{\infty} df\frac{\Omega^2_{GW}}{\Sigma^2_{\Omega}},
\end{equation}
with
\begin{equation}
  \begin{split}
     \Sigma_{\Omega}&=\Sigma_{I}\frac{4\pi ^2 f^3}{3H^2_0}, \\
     \Sigma_{I}&\simeq \frac{20\sqrt{2}}{3 L^2}\left[\left(1
+\left(\frac{0.4\text{mHz}}{f}\right)^2\right) \frac{4 S_I(f)}{(2\pi f)^4}+S_{II}(f)\right]\left[1+\left(\frac{f}{4f_{\ast}/3}\right)^2\right],
  \end{split}
\end{equation}
where $T$ is the observation time, $f_{\ast}=c/(2\pi L)$,
the arm length $L=2.5\times 10^9$m, $\sqrt{S_I}=3\times 10^{-15}$m/s$^2$, $\sqrt{S_{II}}=15$pm for LISA,
$L=3\times 10^9$ m, $\sqrt{S_I}=3\times 10^{-15}$m/s$^2$, $\sqrt{S_{II}}=8$pm for TaiJi,
and $L=\sqrt{3}\times 10^8$ m, $\sqrt{S_I}=1\times 10^{-15}$m/s$^2$, $\sqrt{S_{II}}=1$pm for Tianqin.
Taking one year's observational time,
the signal to noise ratio for induced GWs from the parameter set D
is $\text{SNR}=31004$ in LISA, $\text{SNR}=60987$ in TaiJi and $\text{SNR}=2359$ in TianQin.

It was argued that the energy spectrum of secondary GWs has a log-dependent power index if the primordial power spectrum is narrow \cite{Yuan:2019wwo}.
Therefore, it is interesting to investigate the power index $n$ of $\Omega_{\text{GW}}(f)$ produced in this model
because the parametrization of stochastic GW background $\Omega_{\text{GW}}(f)\sim f^{n}$
is a powerful tool in probing the cosmic history \cite{Kuroyanagi:2018csn}.
For example, for the stochastic GW background from the
coalescence of binary compact stars
such as the mergers of binary black holes, neutron stars or white dwarfs, $n=2/3$ before the peak frequency.
We find that $\Omega_{\text{GW}}(f)$
around the peak frequency can be parameterized
as the broken power law form $\Omega_{\text{GW}}(f)\sim f^{n}$ \cite{Xu:2019bdp,Fu:2019vqc} although the value of $n$ depends on the particular model discussed.
As shown in Fig. \ref{Fig:4},
for the parameter set D, $\Omega_{\mathrm{GW}}\sim f^{2.57}$
for $f<f_c=2.5\times 10^{-3}$ Hz
and $\Omega_{\mathrm{GW}}\sim f^{-2.8}$ for $f>f_c$.
For $f>f_c$, the power index $n=-2.8$ is twice of the spectral index of the power spectrum -1.43 found in section \ref{section:2} since $\Omega_{\text{GW}}\sim P_\zeta^2$.
In the infrared regions with $f\ll f_c$, the log-dependent power index is
$n=3-2/\ln(1.1 f_c/f)$ which is similar to the result
$n=3-2/\ln(f_c/f)$ obtained in \cite{Yuan:2019wwo}.

\section{Conclusions}
\label{section:5}

It is possible that most of DM constitutes of PBHs
if the mass of PBH DM is in the order of $10^{-12}\ M_\odot$.
To produce the order one $Y_{\text{PBH}}$ with this mass, the curvature
power spectrum needs to be in the order of 0.01 at the
scale $k\sim 10^{12}\ \text{Mpc}^{-1}$. This large power spectrum
also generates large secondary GWs at the mHz band
which can be observed by the future space based GW observatory
like LISA, TaiJi and TianQin. However, it is difficult to enhance
the power spectrum for a single canonical field inflation.
By considering non-canonical inflation like G or k inflation,
we find a new mechanism to produce PBH DM and secondary
GWs. In particular, the field dependent kinetic term
$[1-2G(\phi)]X$ can arise from G inflation, k inflation
or general scalar tensor theory of gravity,
and we propose to use the function $G(\phi)$ with a peak at $\phi_r$
to enhance the power spectrum at small scales.
The power spectrum is enhanced around the peak, so
the PBH mass and the frequency of secondary GWs
are determined by the value of $\phi_r$.
In other words, we can adjust the value of $\phi_r$
to get the PBH mass and the frequency of secondary GWs we want.
Away from the peak, $G(\phi)$ is negligible and we recover the usual slow-roll
inflation which is constrained by the CMB observations. Around the peak,
the potential becomes effectively a flat plateau and the slow-roll inflation
transiently turns to ultra slow-roll inflation.

We use the power law potential and the function $G(\phi)=d/(1+|\phi-\phi_r|/c)$ as an example to produce non-negligible
PBH abundances with masses around $36.7 M_\odot$, $10^{-3}M_\odot$ and $10^{-12}M_\odot$,
and secondary GWs with frequencies around 10nHz, $10^{-7}$Hz and mHz.
The PBH DM with the stellar mass of $30M_\odot$ could be the black
holes observed by LIGO and Virgo collaboration. To enhance the power spectrum by seven orders of
magnitude, the parameter $d$ should be in the order of $10^8$. The parameter
$c$ is in the order of $10^{-10}$ so that away from the peak
the function $G(\phi)$ is negligible and the usual slow-roll is guaranteed. Therefore, we don't need to search all values of $c$ and $d$ and to fine tune them to several digits.
We give four parameters sets to show how the
enhancement of the power spectrum at different scales can be achieved. By changing
the parameter $c$ from $9.568\times 10^{-11}$ to $9.54\times 10^{-11}$,
the PBH peak abundances decrease from $Y_{\text{PBH}}^{\text{peak}}=0.001$ to $Y_{\text{PBH}}^{\text{peak}}=7.7\times10^{-5}$,
so the model may accommodate more robust constraints on $Y_{\text{PBH}}$.
The secondary GWs generated by the
model have the characteristic power law behaviour $\Omega_{GW}(f)\sim f^{n}$
and are testable by either PTA or LISA/TaiJi/TianQin observations.

\begin{acknowledgments}
This research was supported in part by the National Natural Science Foundation of
China under Grant No. 11875136 and the Major Program of the National Natural Science
Foundation of China under Grant No. 11690021.
\end{acknowledgments}


%

\end{document}